\documentclass[aps,pre,superscriptaddress,%
 amsmath,amssymb,
reprint,%
]{revtex4-2}
\usepackage{float}
\usepackage{graphicx}
\usepackage{dcolumn}
\usepackage{bm}

\usepackage[T1]{fontenc}
\usepackage[english]{babel}
\usepackage{mathptmx}
\usepackage{xcolor}

\usepackage{pgfplots}
\pgfplotsset{compat=1.15}

\newcommand{\erf}{\text{Erf}}

\newcommand{\be}{\begin{equation}}
\newcommand{\ee}{\end{equation}}
\newcommand{\bea}{\begin{eqnarray}}
\newcommand{\eea}{\end{eqnarray}}

\begin{document}


\title[]{Polymer dynamics under tension: mean first passage time for looping}

\author{Wout Laeremans}
\affiliation{Soft Matter and Biological Physics, Department of Applied Physics and Science Education, and Institute for Complex Molecular Systems,
Eindhoven University of Technology, P.O. Box 513, 5600 MB Eindhoven, Netherlands}
\author{Anne Floor den Ouden}
\affiliation{Laboratory of Physical Chemistry, Department of Chemical Engineering and Chemistry,
Eindhoven University of Technology, P.O. Box 513, 5600 MB Eindhoven, Netherlands}
\author{Jef Hooyberghs}
\affiliation{UHasselt, Faculty of Sciences, Data Science Institute, Theory Lab, Agoralaan, 3590 Diepenbeek, Belgium}
\author{Wouter G. Ellenbroek}
\email{w.g.ellenbroek@tue.nl}
\affiliation{Soft Matter and Biological Physics, Department of Applied Physics and Science Education, and Institute for Complex Molecular Systems,
Eindhoven University of Technology, P.O. Box 513, 5600 MB Eindhoven, Netherlands}

\date{\today}

\begin{abstract}
This study deals with polymer looping, an important process in many chemical and biological systems. We investigate basic questions on the looping dynamics of a polymer under tension using the freely-jointed chain (FJC) model. Previous theoretical approaches to polymer looping under tension have relied on barrier escape methods, which assume local equilibrium, an assumption that may not always hold. As a starting point we use an analytical expression for the equilibrium looping probability as a function of the number of monomers and applied force, predicting an inverse relationship between looping time and looping probability. Using molecular dynamics simulations the predictions of this theoretical approach are validated within the numerical precision achieved. We compare our predictions to those of the barrier escape approach, by way of a calculation of the mean first passage time (MFPT) for the ends of a polymer to cross. For this purpose, we derive the exact free energy landscape, but resulting temporal predictions do not agree with the observed inverse scaling. We conclude that the traditional barrier escape approach does not provide satisfactory  predictions for polymer looping dynamics and that the inverse scaling with looping probability offers a more reliable alternative.
\end{abstract}

\maketitle

\section{Introduction}
Polymer looping or polymer cyclization is of importance in many biological and chemical systems and holds promise for future applications in fields such as material science and soft robotics. One example is loop formation in DNA, necessary for many cellular processes such as transcription, genetic recombination, and regulation of DNA replication~\cite{schleif1992dna, matthews1992dna, saiz2006dna}. Similarly, the polypeptide titin contributes to muscle function by folding and unfolding loops under tension, which helps restore muscle length after stretching~\cite{kellermayer1997folding, freundt2019titin}. Additionally, synthetic biomimetic polymers that incorporate similar looping mechanisms have been developed, offering potential advancements in material science by enhancing properties like elasticity and extensibility~\cite{guan2004modular, chung2014direct, biswas2018patterning}. 

To understand and characterise these looping processes, it is interesting to explore how the looping time scales in the number of monomers and the applied force. Experimental studies, such as those using optical tweezers, have demonstrated that even a small increase in force can significantly impact the looping time. For example, studies on DNA and titin have shown that a piconewton increase in force can extend the looping time by an order of magnitude~\cite{chen2010femtonewton, chen2015dynamics}. In absence of tension, polymer cyclization has been extensively studied theoretically, and remains a very complex and challenging topic~\cite{amitai2017polymer, ye2018dynamic, kappler2019cyclization}. However, the effect of tension on the looping time has only been studied a limited number of times on a theoretical basis~\cite{blumberg2005disruption, shin2012effects}. In this work, we present a significant advancement by demonstrating that, for a freely-jointed chain (FJC) model, the looping time inversely scales with the probability of loop formation. Although such an inverse scaling law has been previously reported in the absence of tension~\cite{jun2003diffusion}, its application under tension has not been explored until now. We derive an analytical expression that accounts for both the number of monomers and the applied force, and we validate our theoretical predictions through molecular dynamics (MD) simulations, achieving excellent agreement.

A previous study on polymer looping under tension employed a barrier escape approach~\cite{shin2012effects}. Such methods, initially introduced by Szabo, Schulten, and Schulten~\cite{szabo1980first}, rely on the assumption of local equilibrium. However, the validity of this assumption for polymer cyclization has been questioned, as the relaxation dynamics of the end-to-end distance might be slow compared to the looping time~\cite{cheng2011failure}. A second obstruction in the context of polymers under tension, is constructing a free energy landscape in a single reaction coordinate. Due to the inherent asymmetry of the problem caused by the external force, this cannot be done exactly. To assess the applicability of the barrier escape approach to polymer looping dynamics, we investigate the mean first passage time (MFPT) for the polymer ends to cross along the force direction. This approach allows us to use the projection of the end-to-end distance along the force as a reaction coordinate, effectively addressing the issue of asymmetry. Utilizing the FJC model, we are able to construct the exact free energy landscape, allowing us to evaluate the barrier escape approach with minimal approximation.

Our findings indicate that the barrier escape approach does not match simulations as well as the inverse scaling with the looping probability that we propose in case of a freely-joined chain model. Thus, our results suggest that the inverse scaling might provide a more reliable description of polymer looping dynamics under tension compared to the previously reported barrier escape results. Consequently, our work offers a new framework for explaining simulation and experimental observations, like DNA cylisation~\cite{chen2010femtonewton, chen2015dynamics, starr2022coarse}.

This article is structured as follows. In Sec.~\ref{sec:polymer_looping}, we derive the looping probability and compare its inverse scaling relation to the looping time with MD simulations. Next, in Sec.~\ref{sec:crossing}, we set up the general framework to calculate the MFPT for the ends of a polymer to cross as a barrier escape problem. In Sec.~\ref{sec:effFx}, we derive an expression for the free energy necessary for the MFPT calculation, which we show to be exact in the thermodynamic limit. Using this free energy, we derive in Sec.~\ref{sec:KGpred} an analytical result for the MFPT in case of the crossing event. In Sec.~\ref{sec:corr}, we re-examine the MFPT, but this time numerically using a free energy that gives a correction beyond the thermodynamic limit, as well as using the exact free energy. Finally, in Sec.~\ref{sec:dis_and_conc}, we give a short discussion and summarize the main conclusion of this work. 


\section{Scaling law for polymer looping time}
\label{sec:polymer_looping}
\begin{figure}[t]
    \centering
    \includegraphics[width=0.8\linewidth]{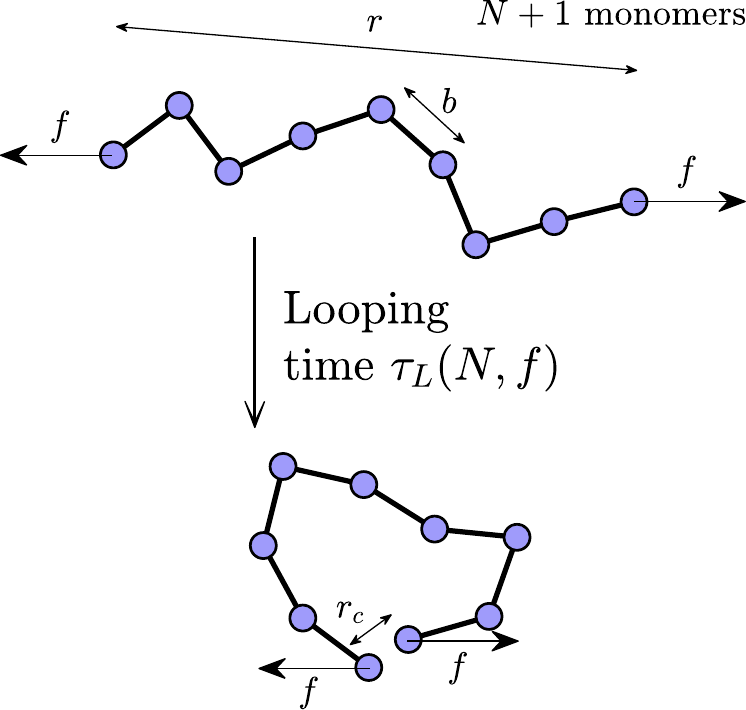}
\caption{Visual representation of the looping process (in three dimensions). Both ends of the polymer are subjected to an external force $f$. The FJC model consists of $N$ joints ($N + 1$ monomers) with bond length $b$ and holds an end-to-end distance $r$. When the ends are as close as the capture radius $r_c$, a loop is formed.}
\label{fig:looping_process} 
\end{figure}

We study a polymer as a collection of $N+1$ beads (monomers), connected by $N$ bonds of length $b$. Then, we define a loop as a configuration (in three dimensions) that has an end-to-end distance $r$ smaller than or equal to a certain capture radius $r_c$, as visualized in Fig.~\ref{fig:looping_process}. The looping time,  denoted by $\tau_L(N,f)$, represents the MFPT over all possible realisations for $r$ starting in equilibrium to become less than or equal to $r_c$, with $f$ the external force on the end-monomers:
\begin{align}
    \tau_L(N,f) &\equiv \left\langle\text{inf}\{t\geq 0: r(t) \leq r_c\} \right\rangle,\label{eq:formdef}
\end{align}
where $\langle\cdot\rangle$ denotes the ensemble average. In general, deriving an expression for $\tau_L(N,f)$ is a highly complex problem, which in case of $f=0$ has been studied repeatedly in the past~\cite{wilemski1974diffusion, wilemski1974diffusion2, doi1975diffusion, szabo1980first, pastor1996diffusion, chen2005diffusion, hyeon2011capturing, toan2008kinetics, amitai2012analysis, afra2015kinetics}. Moreover, it has been noted that the end-to-end distance follows non-Markovian dynamics~\cite{guerin2012non, kappler2019cyclization} and that different backbone models (e.g. Rouse model vs FJC) do not necessarily make the same predictions~\cite{amitai2012analysis}. Nevertheless, we will assume the looping time to be dominated by the slower large length-scale modes (often referred to as diffusion limited~\cite{toan2008kinetics}), leading to~\cite{thirumalai1999time, jun2003role}
\begin{align}
    \tau_L(N,f) &\approx \frac{\tau_0}{P_L(N,f)},
    \label{eq:profass}
\end{align}
with $P_L(N,f)$ the probability to form a loop in equilibrium and $\tau_0$ a reconfiguration time containing the dynamic aspect. As an analogy, one can think of rolling a die with a polymer configuration on every side. Then, the average number of rolls needed to get the looped configuration would be one over the probability of rolling that configuration. Note however, that it would be an unfair die, as every side would have a Boltzmann weight due to the energy introduced by the applied force. Nonetheless, this does not invalidate the reasoning. The factor $\tau_0$ can be thought of as the time it takes to make a single roll. Important is that this analogy only holds when every configuration would be independent, which in turn is only a valid approximation when the time needed to form a loop well exceeds the time needed to get an independent configuration.  

\subsection{Looping probability}
For a FJC, the free energy $F(\vec{r},f)$, with $\vec{r}$ the end-to-end vector and $f$ the external force is given by
\begin{align}
    F(\vec{r}, f) &= -\vec{f}\cdot \vec{r} - TS(\vec{r}) = -fx + F(\vec{r}, f=0),
\end{align}
where the force is taken to be in the $x$-direction. Furthermore, we used that the entropic contribution $TS(\vec{r})$ is simply the free energy at zero force since the FJC contains no energy apart from the $f$-dependent term. From this, one can see that the probability density $P(\vec{r},f)$ becomes
\begin{align}
    P(\vec{r},f) &= \frac{e^{-\beta F(\vec{r},f)}}{Z(f)} = \frac{e^{\beta f x} e^{-\beta F(\vec{r}, f=0)}}{Z(f)} \\
    &= \frac{Z(f=0)}{Z(f)} e^{\beta f x} P(\vec{r}, f=0), \label{eq:Zf}
\end{align}
with $\beta = (k_BT)^{-1}$ the inverse temperature and $Z(f)$ the partition function, given by~\cite{rubinstein2003polymer}
\begin{align}
    Z(f) &= \iiint_{\vec{r}} e^{-\beta \left[ -fx - TS(\vec{r}) \right]} \ d\vec{r} = \left[  \frac{4\pi\sinh\left( \beta f b \right)}{\beta f b} \right]^N. \label{eq:part}
\end{align}
Hence, we arrive at~\cite{dai03}
\begin{align}
    P(\vec{r},f) &= \left[\frac{\sinh(\beta f b)}{\beta f b}\right]^{-N} e^{\beta fx} P(\vec{r},f=0), \label{eq:Pdist}
\end{align}
with $P(\vec{r},f=0)$ approaching a Gaussian probability density~\cite{yamakawa1971modern} for $N\gg 1$ and $r\ll bN$,
\begin{align}
    P(\vec{r}, f=0) &\approx \left(\frac{3}{2\pi N b^2} \right)^{3/2} e^{-\frac{3}{2} \frac{r^2}{Nb^2} }. \label{eq:Gaussian}
\end{align}
\begin{figure}[t]
    \centering
    \includegraphics[width=0.8\linewidth]{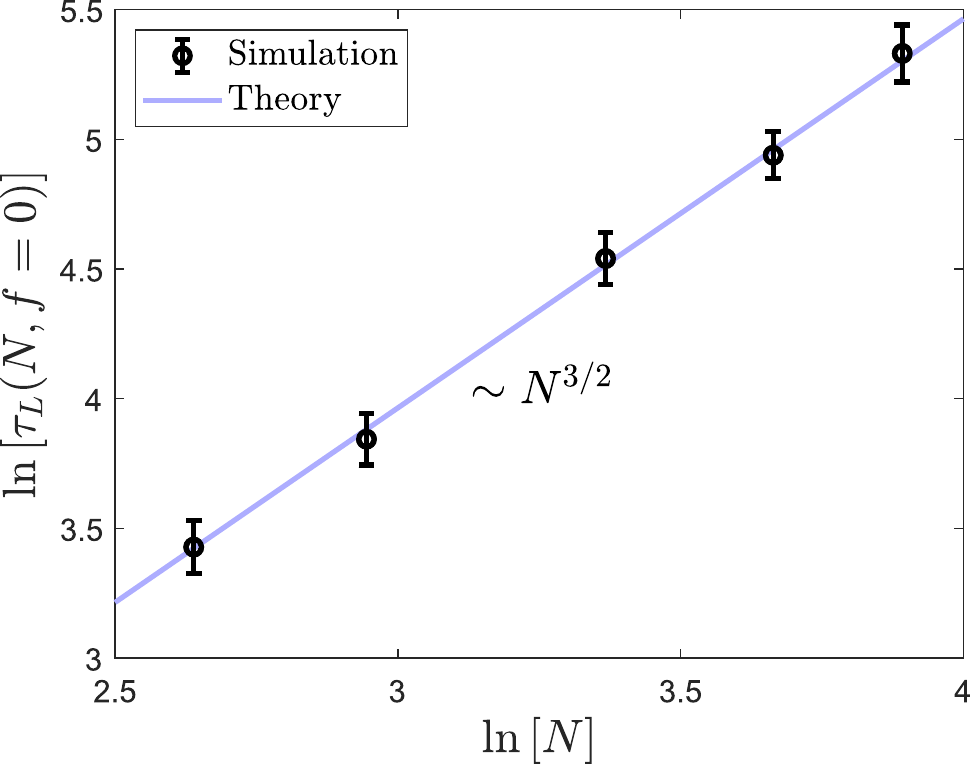}
\caption{Looping time without force. Simulation results of the mean first passage time $\tau_L(N,f=0)$ for a freely-joined chain without tension for a number of joints $N = 14, 19, 29, 39, 49$ with $b = r_c = \beta = \gamma = 1$ in arbitrary units, compared to the $\sim N^{3/2}$ scaling as predicted by Eq.~\ref{eq:Looping_time}. The theory was fitted to the simulation data as a linear fit in a loglog-pace: $\ln[\tau_L(N,f=0)] = \frac{3}{2}\ln[N] + a$, with $a$ the fit parameter.}
\label{fig:tau0_SSS_scaling} 
\end{figure}%
Next, we aim to calculate the looping probability, where we integrate $P(\vec{r},f)$ over a sphere of radius $r_c$. This can be done using spherical coordinates, reading
\begin{align}
    P_L(N,f) &= 2 \pi \left(\frac{3}{2\pi N b^2} \right)^{3/2} \left[\frac{\sinh(\beta f b)}{\beta f b}\right]^{-N} \times \nonumber \\ &\times \int_{0}^{r_c} dr \int_{0}^{\pi} d\theta  \ e^{\beta f r \cos(\theta)} e^{-\frac{3}{2} \frac{r^2}{Nb^2}} r^2 \sin(\theta) \\
    &\sim \alpha^{-3/2}\left[ \frac{\sinh(\beta f b)}{\beta f b} \right]^{-N} e^{ \frac{\beta^2 f^2 \alpha}{4}}  \frac{\alpha}{\beta f} \times \Biggl [  \nonumber \\
    &-2 e^{ \frac{-\left( 2 r_c + \beta f \alpha \right)^2}{4 \alpha}} \left( -1 + e^{2 \beta f r_c} \right) + \nonumber \\
    &+ \beta f\sqrt{\alpha} \left[ \erf\left( \frac{2r_c - \beta f\alpha}{2\sqrt{\alpha}} \right) + \erf\left( \frac{2r_c + \beta f\alpha}{2\sqrt{\alpha}} \right)\right] \Biggr ]. \label{eq:exactIntPL}
\end{align}
Here, we defined $\alpha \equiv 2Nb^2/3$ and the proportionality sign indicates a constant prefactor. For $r_c$ much smaller than the width of the Gaussian term $b\sqrt{N}$ and $\beta f r_c$ not too large, the scaling in $N$ and $f$ is too good approximation given by
\begin{align}
    P_L(N,f) \sim N^{-3/2} \left[\frac{\sinh(\beta f b)}{\beta f b}\right]^{-N},  \label{eq:loopingprob}
\end{align}
which would be obtained by filling out $r = 0$ in Eq.~\ref{eq:Pdist}. This stems from the fact that $P(\vec{r},f)$ is a slowly varying function for $r \leq r_c \ll b\sqrt{N}$.

\subsection{Looping time}
Based on Eq.~\ref{eq:profass} in combination with Eq.~\ref{eq:loopingprob}, we thus predict that the looping time will scale as follows:
\begin{align}
    \tau_L(N,f) &\sim N^{3/2} \left[\frac{\sinh(\beta f b)}{\beta f b}\right]^{N}. \label{eq:Looping_time}
\end{align}
One can note that in the limit of $f \to 0$, the result reduces to the known scaling law $\tau_L(N,f=0) \sim N^{3/2}$, predicted by Szabo, Schulten and Schulten in their so called SSS-theory~\cite{szabo1980first}. This has been observed both in simulations~\cite{pastor1996diffusion, toan2008kinetics,hyeon2011capturing} and in experiments with polypeptides~\cite{bieri1999speed, lapidus2000measuring, sahoo2006temperature}. Including a force, the predicted result of Eq.~\ref{eq:Looping_time} has not been reported nor confirmed to the best of our knowledge. 

\begin{figure}[t]
    \centering
    \includegraphics[width=0.81\linewidth]{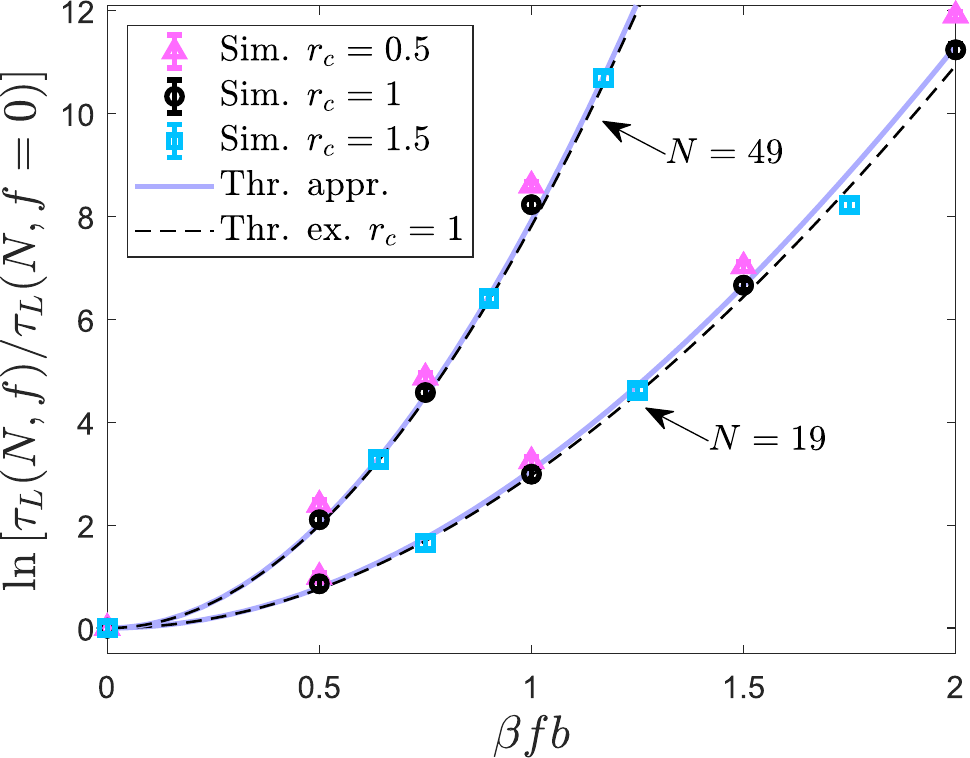}
\caption{Looping time with force. Simulation results of the mean first passage time $\tau_L(N,f)$ for a freely-jointed chain with a number of joints $N = 19$ (lower) and $N = 49$ (upper) and $b = \beta = \gamma = 1$, $r_c = 0.5$ (pink triangle), $r_c = 1$ (black circle) and $r_c = 1.5$ (blue square) in arbitrary units. These are compared to the theoretical prediction of Eq.~\ref{eq:Looping_time} in full purple lines and the exact inverse probability using Eq.~\ref{eq:exactIntPL} for $r_c = 1$. For $r_c$ approaching zero, the exact inverse probability approaches the theory of Eq.~\ref{eq:Looping_time}. The better correspondence between theory and simulation for $r_c = 1.5$ as compared to $r_c = 0.5$ for $N = 49$, stems from the fact that when $r_c \ll b$, contributions from higher modes become more important~\cite{toan2008kinetics}.}
\label{fig:sim} 
\end{figure}

In order to verify the result of Eq.~\ref{eq:Looping_time}, we performed MD simulations. The methodological details of these can be found in Appendix~\ref{app:MD}. First, we checked whether the scaling law $\tau_L(N,f=0) \sim N^{3/2}$ could be recovered. This we did for $N$ ranging from $14$ to $49$, with $r_c = b$. The result, shown in Fig.~\ref{fig:tau0_SSS_scaling}, confirms the correct scaling behavior in $N$. Next, we measured the looping time with an external force applied to the end-monomers. The result is presented in Fig.~\ref{fig:sim}, where the data has been normalized to $\tau_L(N,f=0)$, so $\tau_L(N,f)/\tau_L(N,f=0) = \left[\sinh(\beta f b)/(\beta f b)\right]^{N}$ according to Eq.~\ref{eq:Looping_time}. The full purple lines represent the prediction of Eq.~\ref{eq:Looping_time}, while the dotted black lines give the exact inverse scaling with the looping probability using Eq.~\ref{eq:exactIntPL} for $r_c = 1$. The smaller $r_c$ becomes, the closer the correspondence between both.

Decent agreement between simulation and theory can be observed. However, particularly for $N = 49$, it can be seen that $r_c = 1.5$ aligns better with the theory than $r_c = 0.5$, which is counterintuitive at first glance. This discrepancy can be understood by noting that when $r_c \ll b$, the dynamics of the end-to-end distance is no longer dominated by the global motion of the chain, as previously explained in Ref.~\onlinecite{toan2008kinetics}. The authors highlight that, for a polymer chain to explore the region where $r < b$, faster internal motion becomes necessary, thereby increasing the significance of higher modes in the overall dynamics. Consequently, the prediction of Eq.~\ref{eq:profass} fails for $r_c \ll b$. Nonetheless, this regime is of limited relevance for most practical applications.

\section{Looping as a barrier escape problem}
\label{sec:crossing}
With the analysis in the previous section, excellent results were obtained to predict the scaling of the looping time with respect to the number of joints $N$ and the external force $f$, as shown in Fig.~\ref{fig:tau0_SSS_scaling} and \ref{fig:sim}. Nevertheless, the prediction of Eq.~\ref{eq:profass} does not allow one to extract an actual time as it is not given what $\tau_0$ is. To gain more insight, we will now investigate looping as a barrier escape problem, where we take the viewpoint of one of the ends diffusing in an effective free energy landscape~\cite{szabo1980first}. Then, the MFPT is the time needed for this end to escape its barrier and reach $r = r_c$. However, due to the external force, it is not possible to construct a free energy $F(r)$ in a single reaction coordinate $r$. Therefore, we will study the MFPT for the ends to cross along the direction of the force, making the projection $x$ of the end-to-end distance along the force a suitable reaction coordinate. The crossing time is then defined as: 
\begin{align}
    \tau_{L,x}(N,f) &\equiv \langle \text{inf}\{ t \geq 0 : x(t) \leq 0 \} \rangle. \label{eq:formdefx}
\end{align}
The calculation leading to Eq.~\ref{eq:Pdist} is similar in $x$, leading to~\cite{dai03}
\begin{align}
    P(x,f) &= \left[\frac{\sinh(\beta f b)}{\beta f b}\right]^{-N} e^{\beta fx} P(x,f=0). \label{eq:Pdistxxx}
\end{align}
Therefore, the probability that the ends cross is - under the same approximations as to obtain Eq.~\ref{eq:loopingprob} - given by 
\begin{align}
    P_{L,x}(N,f) &\sim N^{-1/2} \left[\frac{\sinh(\beta f b)}{\beta f b}\right]^{-N}. \label{eq:Ploopx}
\end{align}
If the hypothesis of Eq.~\ref{eq:profass} holds, we thus expect to find that
\begin{align}
    \tau_{L,x}(N,f) &\approx \frac{\tau_{0,x}}{P_{L,x}(N,f)}. \label{eq:profassx}
\end{align}

\begin{figure}[t]
    \centering
    \includegraphics[width=0.8\linewidth]{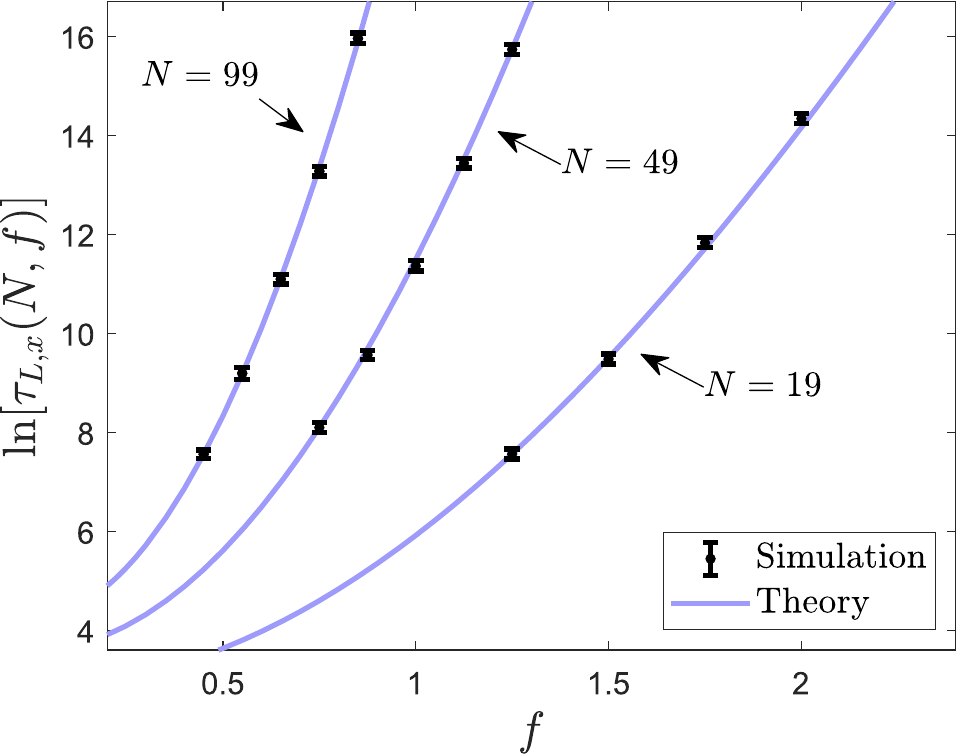}
\caption{Crossing time simulated for different $N$, compared to the theoretical prediction of Eq.~\ref{eq:profassx}. The theoretical prediction is set to match the lowest data point for each $N$. We used $b = \beta = \gamma = 1$ in arbitrary units.} 
\label{fig:checkxscaling} 
\end{figure}

In Fig.~\ref{fig:checkxscaling}, we compared this inverse scaling with MD simulations. Note that we can only perform these simulations at sufficiently high forces, as otherwise the probability of having an initial configuration in which the ends have already crossed is too high and we would obtain $\tau_{L,x} = 0$ for many simulations. Therefore, we cannot rescale at $f = 0$ as was the case for looping. Hence, we took the lowest data point at every $N$ to coincide with the theory. As can be clearly seen in Fig.~\ref{fig:checkxscaling}, the inverse scaling still holds in case of crossing as predicted in Eq.~\ref{eq:profassx}. 

\subsection{MFPT for barrier escape}
The time required for the $x$-coordinates of the two end-monomers of a polymer to cross can be expressed as the MFPT for $x$ to reach zero within an effective free energy landscape $F(x)$. To model this, we consider one of the polymer ends as a particle diffusing within the potential $F(x)$. This scenario involves both a reflecting boundary at $x = L$ (where $L = Nb$ represents the total length of the polymer) and an absorbing boundary at $x = 0$.

The goal is to calculate the time required for the particle, starting in position $x_0$, to reach the boundary $x = 0$ and escape the potential. The probability density $\rho(x,t)$ that describes this system evolves according to the Smoluchowski equation~\cite{risken1996fokker, VANKAMPEN}
\begin{align}
    \partial_t \rho(x,t)  &= -\partial_x \left[ F'(x) \rho(x,t) \right] + D_x \partial_x^2 \rho(x,t),
\end{align}
where $D_x = D/3$, with $D = k_BT/\gamma$ representing the free diffusion coefficient of a single monomer and $\gamma$ denoting the friction coefficient. Calling the time that the particle escapes its barrier $\tau$, one can write~\cite{MFPTbook}
\begin{align}
    \text{Prob}(\tau \geq t) &= \int_{0}^{L} \rho(x', t \ | \ x_0, 0) \ dx' \equiv G(x_0,t),
\end{align}
which is the probability that the particle has remained in the well up to time $t$. As the system is time homogeneous, this probability evolves according to the backward Kolmogorov equation~\cite{risken1996fokker, MFPTbook}
\begin{align}
    \partial_t G(x_0,t) &= F'(x_0) \partial_{x_0} G(x_0,t) + D_x \partial_{x_0}^2 G(x_0,t),
\end{align}
where we impose the boundary conditions~\cite{MFPTbook}
\begin{align}
    \begin{cases}
    \partial_{x_0} G(L,t) &= 0 \qquad \qquad \text{(reflecting boundary)} \\
    G(0,t) &= 0 \qquad \qquad \text{(absorbing boundary)}
    \end{cases}.
\end{align}
Next, the MFPT, meaning going from $x_0$ going to the absorbing boundary set by zero, can be calculated as~\cite{MFPTbook}
\begin{align}
    \tau_{L,x}(N,f; x_0 \rightarrow 0) &= - \int_{0}^{\infty} t \partial_t G(x_0,t) \ dt = \int_{0}^{\infty} G(x_0,t) \ dt.
\end{align}
With the necessary physics now fully explained, the remaining task is merely mathematical. The calculations can be found in Ref.~\onlinecite{MFPTbook}, leading to~\cite{MFPTbook, shin2012effects}.
\begin{align}
    \tau_{L,x}(N,f; x_0 \rightarrow 0) &= \frac{1}{D_{x}} \int_{0}^{x_0}  dx \int_{x}^{L}  dx' \ e^{\beta \left[ F(x)- F(x') \right]} .
\end{align}
Finally, we note that the initial position $x = x_0$ is not fixed, but distributed according to the Boltzmann distribution. To account for this, we average over the initial position $\langle \tau_{L,x}(N,f;x_0 \rightarrow 0) \rangle_{x_0} = 1/Z \int_{0}^{L} \tau_{L,x}(N,f;x_0 \rightarrow 0) e^{-\beta F(x'')} \ dx''$, which results in the final form of the MFPT~\cite{shin2012effects}
\begin{align}
    \tau_{L,x}(N,f) &= \frac{1}{D_{x} Z} \int_{0}^{L}  dx  \int_{0}^{x}  dx'  \int_{x'}^{L}  dx'' \ e^{-\beta \left[ F(x) - F(x') + F(x'') \right]}, \label{eq:tripple_integral}
\end{align}
with $Z$ the partition function 
\begin{align}
    Z &= \int_{0}^{L} e^{-\beta F(x)} \ dx.
\end{align}
For $f$ large enough, one can assume however the starting point to be simply the minimum of the free energy landscape $x = \langle x \rangle$, by which the averaging over the initial position can be neglected
\begin{align}
    \tau_{L,x}(N,f) &\approx \frac{1}{D_{x}} \int_{0}^{\langle x \rangle}  dx  \int_{x}^{L}  dx' \ e^{\beta \left[ F(x)- F(x') \right]}. \label{eq:FP}
\end{align}

\begin{figure}[t]
    \centering
    \includegraphics[width=0.9\linewidth]{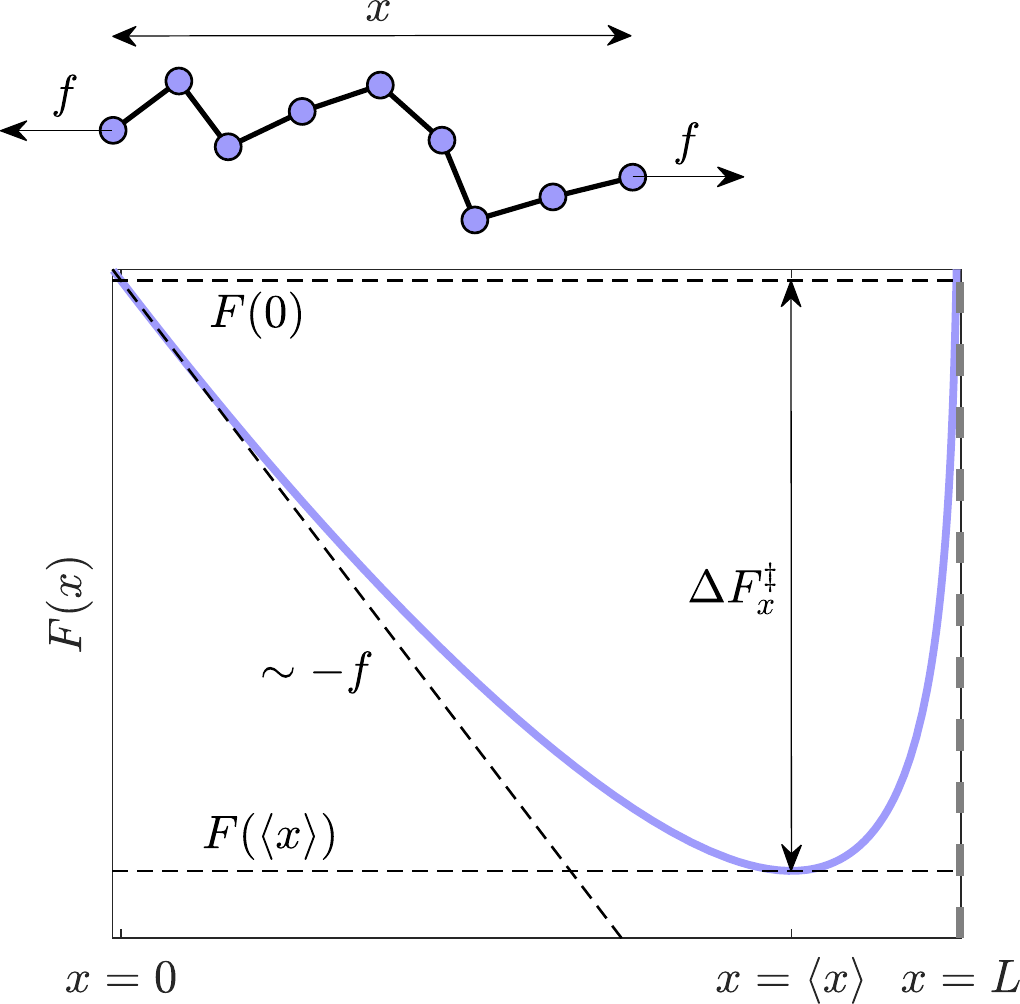}
\caption{Effective free energy $F(x)$ for a freely-jointed chain in function of the projection $x$ of the end-to-end distance along the direction of the force $f$.}
\label{fig:Gen_pol_Fx} 
\end{figure}

When the free energy barrier holds the shape of a well followed by a hill, this double integral is well approximated by Kramers theory~\cite{kramers1940brownian, michaels2018reaction}. Different shapes for the free energy landscape have been worked out in the past as well, e.g. the linear-cubic and cusp~\cite{dudko2006intrinsic, bullerjahn2016analytical}. We propose a new approximation for Eq.~\ref{eq:FP} using similar approaches, but specifically designed for the free energy of a FJC under tension. In Fig.~\ref{fig:Gen_pol_Fx} we show this typical shape for $F(x)$. At small $x$, the shape is linear with a slope of $-f$. This is because the work done by the external force is the main contribution to keep the end-monomers apart at small separations. Therefore, $F(x \approx 0) \approx F(0) - fx$ close to $x = 0$. Around the minimum of the free energy $\left[\langle x \rangle, F(\langle x \rangle)\right]$, the shape is fairly quadratic: $F(x \approx \langle x \rangle) \approx F(\langle x \rangle) + F''(\langle x \rangle) (x - \langle x \rangle)^2/2$. Denoting the free energy difference between this of a crossed configuration ($x = 0$) and the average end-to-end distance ($x = \langle x \rangle$) as $\Delta F_x^\ddagger(f) \equiv F(0) - F(\langle x \rangle)$, one can show that at sufficiently large forces (see Appendix~\ref{app:SP})
\begin{align}
    \tau_{L,x}(N,f) &\approx \frac{1}{D_{x} \beta f} \sqrt{\frac{2\pi}{\beta F''(\langle x \rangle)}} \times e^{\beta \Delta F_x^\ddagger(f)}. \label{eq:SP}
\end{align}
Note that $\langle x \rangle$ depends on $f$, hence the second derivative evaluated here does so implicitly as well.  

We note however, that a FJC chain of $N+1$ monomers is described by $N-1$ stochastic variables, namely the angles in between the joints. In a barrier escape theory, similar to SSS-theory~\cite{szabo1980first}, this high dimensional problem is reduced to a single coordinate description, the end-to-end projection $x$. This coarse-grained picture is only valid when all degrees of freedom have fully relaxed. The more the system is dominated by entropy, the less likely this will be~\cite{cheng2011failure}. For more stiff polymers, this relaxation will be faster, making a single coordinate theory more reliable~\cite{afra2015kinetics}. Hence, it is not guaranteed that a barrier-based model will provide good results, especially in the case of a FJC.

\section{Effective free energy}
\label{sec:effFx}
In order to evaluate the MFPT, either with the full result (Eq.~\ref{eq:tripple_integral}) or with the approximation (Eq.~\ref{eq:SP}), we need to construct a free energy $F(x)$.
Using the isotensional (often referred to as Gibbs~\cite{manca2014equivalence}) free energy $\beta G(f) = - \ln\left[Z(f)\right]$, with $Z(f)$ given by Eq.~\ref{eq:part}, the force-extension relation can be found as $\langle x \rangle = - dG(f)/df$. This results in the following relation as originally derived by Kuhn and Grün~\cite{kuhn1942beziehungen, rubinstein2003polymer}
\begin{align}
    \langle x \rangle &= L\mathcal{L}(\beta f b), \label{eq:force_extension}
\end{align}
with $\langle x \rangle$ the average extension under the applied force $f$ and $\mathcal{L}(y) = \frac{1}{\tanh(y)} - \frac{1}{y}$ the Langevin function. The entropic force is found by inversion of the force-extension relation
\begin{align}
    f_\text{entr.} = \frac{1}{\beta b} \mathcal{L}^{-1}\left( \frac{x}{L} \right), \label{eq:entropic_force}
\end{align}
as at $x = \langle x \rangle$ it should cancel the external force. As the derivative of a thermodynamic potential yields a generalized force with negative sign, we have the derivative of the free energy in $x$
\begin{align}
    F'(x) &= \frac{1}{\beta b} \mathcal{L}^{-1}\left( \frac{x}{L} \right) - f. \label{eq:first_der}
\end{align}
From this, the free energy landscape can be found as
\begin{align}
    &\beta F(x) = \beta \int_0^x F'(\tilde{x}) \ d\tilde{x} \\
 &= -N\left[\ln\left( \frac{\sinh\left[\mathcal{L}^{-1}\left(\frac{x}{L}\right)\right]}{\mathcal{L}^{-1}\left(\frac{x}{L}\right)} \right) - \frac{x}{L}\mathcal{L}^{-1}\left( \frac{x}{L} \right)\right] - \beta fx, \label{eq:F_full_FJC}
\end{align}
where we used that $\int g^{-1}(y) \ dy = yg^{-1}(y) - G \circ g^{-1}(y) + C$, with $g(y)$ a function in $y$, $G$ its antiderivative and $C$ an integration constant. 

\subsection{Validity and thermodynamic limit}
We could have also derived the result of Eq.~\ref{eq:F_full_FJC} more formally. Denoting a microstate as $\omega$, we write the projection of the end-to-end distance along the force for a specific microstate $\omega$ as
\begin{align}
    x(\omega) = \sum_{i=1}^N \Vec{r}_i \cdot \hat{x},
\end{align}
with $\vec{r}_i$ the position vector of monomer $i$. Next, we introduce the subspace volume $\Omega(x)$ associated with a specific $x$-value:
\begin{equation}
    \Omega(x) = \int_{\Lambda} \delta\left(x(\omega) - x \right) \ d\omega,
\end{equation}
of which the total volume is $|\Lambda| = \left(4\pi b^2 \right)^N$. From this, we identify the entropy
\begin{align}
    S(x) = k_B \ln\left[ \Omega(x) \right].
\end{align}
This entropy can be linked to the partition function
\begin{align}
    Z(f) = \int_{\mathbb{R}} e^{-\beta (-fx - T S)} \ dx = \left[  \frac{4\pi\sinh\left( \beta f b \right)}{\beta f b} \right]^N,
\end{align}
by means of the Legendre transform~\cite{callen1960thermodynamics}
\begin{align}
    \ln\left[ \Omega(x) \right] &= \ln\left( Z\left[f(x) \right] \right) - \beta f x.
\end{align}
Filling these out, we get
\begin{align}
    \ln\left[ \Omega(x) \right] &= N\ln\left( 4\pi   \right)  \nonumber \\ &+N\left[\ln\left( \frac{\sinh\left[\mathcal{L}^{-1}\left(\frac{x}{L}\right)\right]}{\mathcal{L}^{-1}\left(\frac{x}{L}\right)} \right) - \frac{x}{L}\mathcal{L}^{-1}\left( \frac{x}{L} \right)\right].
\end{align}
Dropping the constant term and using $F(x) = -fx - TS(x)$, we finally arrive at what was given in Eq.~\ref{eq:F_full_FJC}. From this more formal derivation, it is clear that the entropic term of $F(x)$ is the Legendre transform of $G(f)$. Hence, $F(x)$ is only exact in the thermodynamic limit.

\section{Calculation of the MFPT}
\label{sec:KGpred}
In this section, we will evaluate the MFPT as predicted by Eq.~\ref{eq:SP} for the ends of the polymer to cross along the direction of the force, using the free energy of Eq.~\ref{eq:F_full_FJC}. At the minimum of the free energy, the total or generalised force $f - f_\text{entr.}$ is zero (see Eq.~\ref{eq:first_der}), meaning that the two final terms in Eq.~\ref{eq:F_full_FJC} cancel, so one is left with
\begin{align}
    \beta F(\langle x \rangle) &= -N\ln \left( \frac{\sinh\left[\mathcal{L}^{-1}\left(\frac{\langle x \rangle}{L}\right)\right]}{\mathcal{L}^{-1}\left(\frac{\langle x \rangle}{L}\right)} \right).
\end{align}
Using that $F(0) = 0$ and invoking the force-extension relation $\langle x \rangle = L\mathcal{L}(\beta f b)$ then results in
\begin{align}
    \beta \Delta F_x^\ddagger(f) &= N\ln \left[ \frac{\sinh(\beta f b)}{\beta f b} \right].  \label{eq:DFxxx}
\end{align}
One can note that this is simply $\beta G(f)$, as in the thermodynamic limit, the minimum of the free energy $F(x)$ is the isotensional free energy $G(f)$. Next, the second derivative of the free energy can be calculated by taking the derivative of Eq.~\ref{eq:first_der}. Using that the derivative of the inverse of a function $g$ evaluated in a point $a$ is given by: $\left( g^{-1} \right)'(a) = \frac{1}{g'\left( g^{-1}(a) \right)}$, this can be evaluated in an exact form, leading to

\begin{align}
    F''(\langle x \rangle) &= \frac{1}{\beta b L \frac{d}{d(\beta f b)} \left[ \mathcal{L}\left( \beta f b
 \right) \right]}  \\
 &= \frac{1}{\beta b L \left[ \frac{1}{(\beta f b )^2} - \frac{1}{\sinh^2\left( \beta f b \right)} \right]}. \label{eq:secder}
\end{align}
Plugging in these results in Eq.~\ref{eq:SP}, we find
\begin{align}
    &\tau_{L,x}(N,f) \nonumber \\ &\approx \frac{\sqrt{2\pi bL}}{D_x\beta f} \sqrt{\frac{1}{(\beta fb )^2} - \frac{1}{\sinh^2\left( \beta f b \right)}} \left[ \frac{\sinh(\beta f b)}{\beta f b} \right]^N. \label{eq:SPworkedout}
\end{align}
For $\beta f b \gg 1$, this becomes
\begin{align}
    \tau_{L,x}(N,\beta f b \gg 1) \sim \frac{N^{1/2}}{(\beta f b)^2} \left[\frac{\sinh(\beta f b)}{\beta f b} \right]^N, \label{eq:SSS_f2}
\end{align}
which does not agree with the hypothesis of Eq.~\ref{eq:profassx}, as we identify $\tau_{0,x} \sim f^{-2}$ instead of a constant reconfiguration time. 

\begin{figure}[t]
    \centering
    \includegraphics[width=0.8\linewidth]{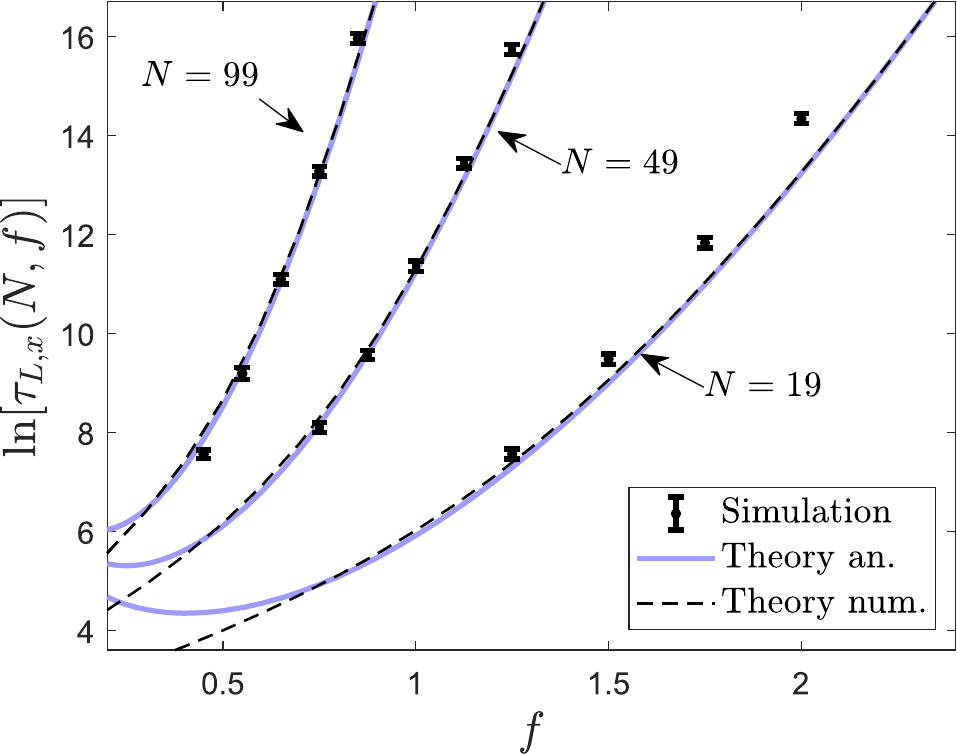}
\caption{Crossing time simulated for different $N$, compared to theoretical prediction of Eq.~\ref{eq:SPworkedout} (full purple lines). The dotted black lines represent the numerical integration of Eq.~\ref{eq:tripple_integral} for the free energy of Eq.~\ref{eq:F_full_FJC}. We used $b = \beta = \gamma = 1$ in arbitrary units.}
\label{fig:theoryvsnumint_TL} 
\end{figure}

In Fig.~\ref{fig:theoryvsnumint_TL}, we show the theoretical prediction of Eq.~\ref{eq:SPworkedout} (full line), together with a numerical integration of Eq.~\ref{eq:tripple_integral} for the free energy of Eq.~\ref{eq:F_full_FJC} (dotted line). For the numerical integration, we replaced the inverse Langevin function by the Padé approximation $\mathcal{L}^{-1}(y) \approx y\frac{3 - y^2}{1 - y^2}$, which describes the inverse Langevin function well on the whole domain~\cite{cohen1991pade, jedynak2015approximation}. From this, it is clear that the analytical result agrees with numerical integration at sufficiently high forces, confirming the validity of Eq.~\ref{eq:SP}. Although our theoretical prediction does not perfectly describe the simulation data, it does give a decent estimate and predicts the right order of magnitude. 

Comparing however the result from the barrier escape approach derived in this section, with the inverse scaling with the looping probability visualized before in Fig.~\ref{fig:checkxscaling}, it becomes clear that the former yields a less accurate prediction. Nevertheless, as we pointed out in Sec.~\ref{sec:effFx}, the free energy that we used is not exact, which is why we cannot exclude the possibility that the poorer prediction might stem from this approximation. To address this uncertainty, we will re-examine the MFPT numerically in the next section, but using a corrected free energy.

\section{Corrected free energy}
\label{sec:corr}
As explained in the derivation of the free energy, it is only exact in the thermodynamic limit. We can, however, improve on this by defining the free energy as
\begin{align}
    \beta F(x) &= -\ln\left[ P(x,f=0) \right] - \beta f x,
\end{align}
and using a probability density $P(x,f=0)$ that offers a higher degree of accuracy. The exact distribution is known, valid for any $N$ in the whole domain $-L \leq x \leq L$, and reads~\cite{dai03}
\begin{align}
    P_\text{ex.}(x, f=0) &= \frac{1}{2^Nb(N-2)!} \int_{x/b}^{N} \sum_{k=0}^{(N-y)/2} (-1)^N  \times \nonumber \\ &\times \binom{k}{N}(N - 2k - y)^{N-2} \ dy \qquad , \quad x \geq 0. \label{eq:dist_ex_distsec}
\end{align}
This probability density can also be rewritten without an integral, which is usually easier and faster to use for numerical calculation. As we show in Appendix~\ref{app:We}, this results in
\begin{align}
    P_\text{ex.}(x, f=0) &= \frac{1}{2^Nb(N-1)!} \sum_{k=0}^N \binom{N}{k} (-1)^k \times \nonumber \\ 
    &\times (N - x/b - 2k)^{N-1} \Theta(N - x/b - 2k), \label{eq:WE}
\end{align}
with $\Theta$ the Heaviside step function.
Nevertheless, it remains a burden to evaluate this expression in Eq.~\ref{eq:tripple_integral}, especially when $N$ is large. In this work, we managed to derive a new expression for $P(x, f=0)$, valid for $N$ large, without any restriction on $x$. The analysis is very similar to the approach followed in Ref.~\onlinecite{yamakawa1971modern} to derive an expression $P(r,f=0)$ in terms of the end-to-end distance $r$. Our calculation in $x$ leads to (see Appendix~\ref{app:JH})
\begin{align}
    P_\text{JH}(x,f=0) &= \frac{\left( 2\pi N b^2 \right)^{-1/2}}{\sqrt{\left[\frac{1}{\mathcal{L}^{-1}(x/L)}\right]^2 - \left[\frac{1}{\sinh\left( \mathcal{L}^{-1}(x/L) \right)}\right]^2}} \times \nonumber \\
    &\times \left[ \frac{\sinh\left( \mathcal{L}^{-1}(x/L) \right)}{\mathcal{L}^{-1}(x/L) e^{x/L\mathcal{L}^{-1}(x/L)}} \right]^N. \label{eq:JH}
\end{align}

\begin{figure}[t]
    \centering
    \includegraphics[width=0.8\linewidth]{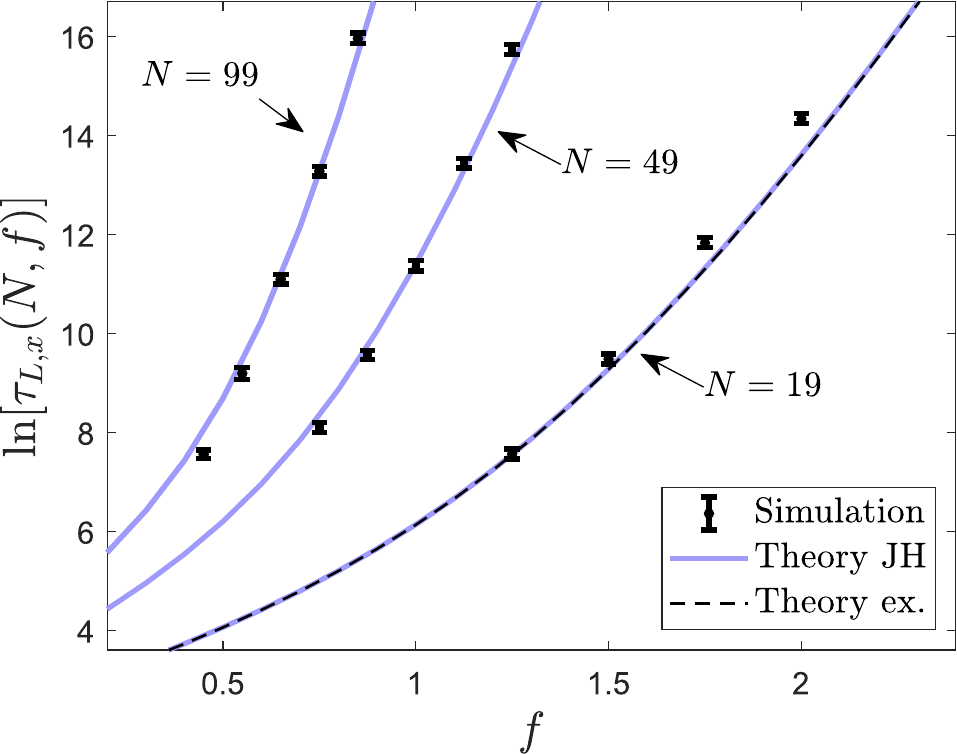}
\caption{Crossing time simulated for different $N$, compared to the numerical integration of Eq.~\ref{eq:tripple_integral}. For the free energy, we used $\beta F(x) = -\ln\left[ P(x,f=0) \right] - \beta fx$, where the full purple lines correspond to $P(x,f=0) = P_\text{JH}(x,f=0)$ (Eq.~\ref{eq:JH}), while the dotted black line corresponds to $P(x,f=0) = P_\text{ex.}(x,f=0)$ (Eq.~\ref{eq:WE}). We used $b = \beta = \gamma = 1$ in arbitrary units.}
\label{fig:num_int_JH_vs_ex} 
\end{figure}
\subsection{Numerical evaluation of the MFPT}
In Fig.~\ref{fig:num_int_JH_vs_ex}, we show the numerical integration of Eq.~\ref{eq:tripple_integral}, using the free energy $\beta F(x) = -\ln\left[ P(x,f=0) \right] - \beta f x$ with $P(x,f=0) = P_\text{JH}(x,f=0)$ as full purple lines. The dotted black line is the numerical integration using $P(x,f=0) = P_\text{ex.}(x,f=0)$. The latter we only solved for $N = 19$, as it is quite time consuming. However, it is clear that there is no noticeable difference between the integrations using $P_\text{JH}$ and $P_\text{ex.}$. As $P_\text{JH}$ was dervied for large $N$, the similarity is expected to even improve for larger $N$.

Although we see an improvement over the analytical calculation using the free energy of Eq.~\ref{eq:F_full_FJC} (see Fig.~\ref{fig:theoryvsnumint_TL}), the theoretical curve still does not match the simulation data as well as the inverse scaling with the looping probability (Fig.~\ref{fig:checkxscaling}). That is, even with the exact free energy, the barrier escape approach provides a less accurate prediction compared to the inverse scaling with the looping probability. This leads us to conclude that the mismatch between theory and simulation in case of the barrier escape approach is intrinsic to the approach itself rather than due to approximations made in its evaluation. This conclusion aligns with the discussion at the end of Sec.~\ref{sec:crossing}, where we explained that single coordinate theories might not produce reliable results due to the local equilibrium assumption. Nonetheless, the barrier escape approach does provide a reasonable estimate for the crossing time without the need for any free parameters.

\section{Discussion and Conclusion}
\label{sec:dis_and_conc}
In this work, we demonstrated that the looping time of a freely-jointed chain (FJC) under tension exhibits an inverse scaling with the equilibrium probability of loop formation. Although such inverse scaling relationships have been proposed previously~\cite{szabo1980first, jun2003diffusion}, they did not account for the presence of an applied force. In our analysis, we explicitly derived how the looping probability depends on both the number of monomers and the applied force, culminating in Eq.~\ref{eq:Looping_time} as our prediction for the looping time. This prediction shows excellent agreement with molecular dynamics simulations, as illustrated in Fig.~\ref{fig:tau0_SSS_scaling} and ~\ref{fig:sim}.

The looping time of a polymer under tension has been previously investigated theoretically using a barrier escape approach~\cite{shin2012effects}. This method relies on the assumption of local equilibrium, where all the degrees of freedom between consecutive joints are reduced to a single reaction coordinate. However, the validity of this assumption is not always clear~\cite{cheng2011failure}, considering that the relaxation dynamics of the end-to-end distance may be slow in comparison to the looping time. A second challenge lies in constructing a free energy in a single coordinate, as the application of a force introduces asymmetry. To address this, we investigated the mean first passage time (MFPT) for the ends of the polymer to cross along the direction of the force, employing the barrier escape approach. This allowed to consider the projection of the end-to-end distance along the force direction as a suitable reaction coordinate, which avoids any problem with asymmetry. Using an approximate free energy, we derived an analytical expression for the MFPT (Eq.~\ref{eq:SPworkedout}). However, this result did not align with our previous conclusion that the looping time should scale inversely with the looping probability. A key advantage of working with the FJC model, is that it provides access to the exact free energy. To determine whether the disagreement was due to the approximations made in the free energy, we also numerically investigated the MFPT using the exact free energy. As demonstrated in Fig.~\ref{fig:num_int_JH_vs_ex}, this numerical result also failed to exhibit the expected inverse scaling with the looping probability. Thus, we present compelling evidence that the barrier escape approach, in the context of polymer looping, does not always yield reliable predictions.

Although we conducted all calculations using the FJC model, which facilitates both analytical and exact calculations, we anticipate that the findings of our work will be broadly applicable across a wide range of systems. For instance, recent work presented in Ref.~\onlinecite{starr2022coarse} investigated the effects of supercoiling and loop length on DNA cyclization. In that study, the authors employed the barrier escape approach outlined in Eq.~\ref{eq:tripple_integral} to determine the MFPT for loop formation, with the free energy landscape derived from simulations. Our results suggest that considering the inverse scaling with the looping probability might yield more reliable predictions in such systems.


\acknowledgments{This research is financially supported by the Dutch Ministry of Education, Culture and Science (Gravity Program 024.005.020 – Interactive Polymer Materials IPM).
}

\bibliography{references}

\appendix

\section{Derivation to go from Eq.~\ref{eq:FP} to Eq.~\ref{eq:SP}}
\label{app:SP}
\begin{figure}[t]
    \centering
    \includegraphics[width=0.8\linewidth]{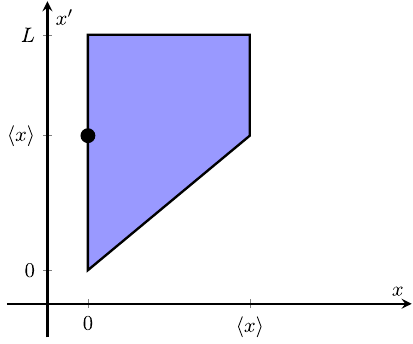}
\caption{Integration domain for Eq.~\ref{apeq:start} colored purple. The integrand is maximised at the boundary of the domain at $[x,x'] = [0, \langle x \rangle]$, given by a black dot.}
\label{fig:Fig_Appendix} 
\end{figure}

Here, we will derive the transition from Eq.~\ref{eq:FP} to Eq.~\ref{eq:SP}, very similar to the work of Kramer~\cite{kramers1940brownian}. We start by noting that Eq.~\ref{eq:FP} can be split up into two parts
\begin{align}
    \tau_{L,x}(N,f) &= \frac{1}{D_{x}} \int_{0}^{\langle x \rangle} e^{\beta F(x)} \left( \int_{x}^L e^{-\beta F(x')} \ dx' \right) \ dx. \label{apeq:start}
\end{align}
The integration domain of the inner part always contains $x' = \langle x\rangle$, which is the minimum of the free energy landscape (see Fig.~\ref{fig:Fig_Appendix}). Therefore, one can approximate
\begin{align}
    F(x' \approx \langle x \rangle) \approx F(\langle x \rangle) + \frac{F''(\langle x \rangle)}{2} (x' - \langle x \rangle)^2.
\end{align}
As the main contribution comes from the minimum, this leads to
\begin{align}
    &\tau_{L,x}(N,f) \nonumber \\ &\approx \frac{1}{D_{x}} \int_{0}^{\langle x \rangle} e^{\beta F(x)} \left( \int_{x}^L e^{-\beta\left[ F(\langle x \rangle) + \frac{F''(\langle x \rangle)}{2} (x' - \langle x \rangle)^2 \right]} \ dx' \right) \ dx.
\end{align}

To good approximation, the boundaries of integration of the inner part can be taken to infinity, as the integrand rapidly decreases around $x' = \langle x \rangle$, leading to
\begin{align}
    &\tau_{L,x}(N,f) \nonumber \\ &\approx \frac{1}{D_{x}} \int_{0}^{\langle x \rangle} e^{\beta F(x)} \left( \int_{-\infty}^{+\infty} e^{-\beta\left[ F(\langle x \rangle) + \frac{F''(\langle x \rangle)}{2} (x' - \langle x \rangle)^2 \right]} \ dx' \right) \ dx \\
    &= \frac{1}{D_{x}} \sqrt{\frac{2\pi}{\beta F''(\langle x \rangle)}} e^{- \beta F(\langle x \rangle)}   \int_{0}^{\langle x \rangle} e^{\beta F(x)} \ dx.
\end{align}

The main contribution to the final integral comes from the boundary $x = 0$. As this is a very unfavourable state, the monomers want to be pulled apart, making a linear approximation valid around this point with of slope of $-f$ (see Fig.~\ref{fig:Gen_pol_Fx}):
\begin{align}
    F(x \approx 0) \approx F(0) - fx.
\end{align}
As contributions to larger $x$ become small quickly, the right integration boundary can be taken to infinity again, giving
\begin{align}
    \int_{0}^{\langle x \rangle} e^{\beta F(x)} \ dx &\approx \int_{0}^{+\infty} e^{\beta\left[ F(0) - fx\right] } \ dx= \frac{e^{\beta F(0)}}{\beta f}.
\end{align}
This in turn leads to
\begin{align}
    \tau_{L,x}(N,f) &\approx \frac{1}{D_{x} \beta f} \sqrt{\frac{2\pi}{\beta F''(\langle x \rangle)}} \times e^{\beta \left[F(0) - F(\langle x \rangle) \right]} \\
    &= \frac{1}{D_{x} \beta f} \sqrt{\frac{2\pi}{\beta F''(\langle x \rangle)}} \times e^{\beta \Delta F_x^\ddagger(f)}, \label{eqap:tauapprox}
\end{align}
with $\Delta F_x^\ddagger = F(0) - F(\langle x \rangle) = \text{max} \{ F(x) - F(x') \quad , \ x \leq x' \leq L \ \land \ 0 \leq x \leq \langle x \rangle \}$, as was stated in Eq.~\ref{eq:SP} of the main text. Note however, that the approximations used are only valid for $f$ sufficiently large.

\section{MD simulations}
\label{app:MD}
All simulations were performed using the LAMMPS MD simulator~\cite{LAMMPS}. The input parameters below are reported in Lennard-Jones (LJ) dimensionless units. We simulated the FJC as a bead spring model with energy
\begin{align}
    U_S &= \frac{K}{2}\sum_{i = 0}^{N-1} \left( |\vec{r}_{i+1} - \vec{r}_{i}| - b \right)^2,
\end{align}
where the spring constant was taken to be $K = 100$ and the rest length $b = 1$. The value of the spring constant is large enough, such that fluctuations around the rest length are only $\sim10\%$. Before each measurement, all beads (having mass $m = 1$) were given a random velocity sampled from a Gaussian distribution with variance 1 and an equilibrium run was performed for $10^5$ timesteps, where the timestep was chosen to be $\Delta t = 0.05$. During this equilibration process, we implemented Langevin dynamics for all the monomers using \textit{fix langevin}. After equilibration, we lowered the timestep to $\Delta t = 0.005$ and we used \textit{fix ffl} for the dynamics instead~\cite{hijazi2018fast}, with the friction coefficient $\gamma = 1$. Every data point in the main text is an average over 100 simulations with a different seed. To obtain the MFPT in case of looping, we used the LAMMPS package REACTER~\cite{gissinger2020reacter}, where we defined a reaction for the two end-monomers of the polymer. This allowed for an efficient way of keeping track when the ends are in close contact ($r \leq r_c$). 

\subsection{Error analysis}

Each data point is an average over 100 simulations. Denoting the looping time of a single simulation as $\tau_L^i(N,f)$, the average looping time is given by
\begin{align}
    \tau_L(N,f) &= \frac{1}{100}\sum_{i=1}^{100} \tau_L^i(N,f)
\end{align}
and the standard error of the mean (SEM) is calculated as
\begin{align}
    \text{SEM}\left[ \tau(N,f) \right] &= \frac{\sigma\left[ \tau(N,f) \right]}{\sqrt{100}}, 
\end{align}
where $\sigma$ denotes the standard deviation. The error in the natural logarithm of the average looping time is then given by
\begin{align}
    \Delta \ln\left[ \tau_L(N,f) \right]  &= \frac{\text{SEM}\left[ \tau(N,f) \right]}{\tau_L(N,f)}.
\end{align}
In Fig.~\ref{fig:sim}, we rescaled by $\tau_L(N,f=0)$, a constant for each $N$. Hence, the rescaling does not alter the error in the natural log.

\section{JH probability distribution}
\label{app:JH}
In order to derive Eq.~\ref{eq:JH}, we closely follow the saddle-point or steepest descent approach in Ref.~\onlinecite{yamakawa1971modern}, but modified for $x$ instead of
$r$. Denoting the bond probability $\rho(x_j)dx_j$ as the probability that the joint $j$ has as a projection along the $x$-axis in $[x_j, x_j + dx_j]$, we have
\begin{align}
    \rho(x_i) &= \frac{1}{2b} \Theta(b - |x_i|),
\end{align}
which simply states it is uniform as we expect for a FJC or random walk. Calling now $x = \sum_i x_i$, we have
\begin{align}
    P(x) &= \int_{\{x_n\}} \delta\left(x - \sum_j x_j\right) \prod_{j=1}^N \rho(x_j) \ dx_j. \label{eq:Pjhalsodensity}
\end{align}
Next, we rewrite the delta function as
\begin{align}
    \delta(x) &= \frac{1}{2\pi} \int_{-\infty}^{+\infty} e^{ikx} dk, \label{eq:FourierDelta}
\end{align}
to obtain 
\begin{align}
    P(x) &= \frac{1}{2\pi} \int_{-\infty}^{+\infty} K(k) e^{-ikx} dk,
\end{align}
with 
\begin{align}
    K(k) &\equiv \prod_{j=1}^N \int_{-b}^{b} \rho(x_j) e^{ikx_j} dx_j \\
    &= \prod_{j=1}^N \frac{1}{2b} \int_{-b}^{b} \left[ \cos(kx_j) + i\sin(kx_j) \right] dx_j \\
    &= \left[\frac{\sin(kb)}{kb}\right]^N.
\end{align}
Hence, we find
\begin{align}
    P(x) &= \frac{1}{2\pi} \int_{-\infty}^{+\infty} \left[ \frac{\sin(kb)}{kb} \right]^N e^{-ikx} \ dk \\
    &= \frac{1}{2\pi} \int_{-\infty}^{+\infty} \left[ \frac{\sin(kb)}{kb} \right]^N e^{ikx} \ dk,
\end{align}
where in the last step we used that $\sin(y)/y$ is even. Next, we can use $bk$ as the integration variable, giving
\begin{align}
    P(x) &= \frac{1}{2\pi b} \int_{-\infty}^{+\infty} \left[ \frac{\sin(kb)}{kb} \right]^N e^{iNbk\frac{x}{L}} \ d(bk) \\
    &= \frac{1}{2\pi b} \int_{-\infty}^{+\infty} e^{N\left( \ln\left[ \sin(kb)/(kb) \right] + ikbx/L \right)} \ d(bk) \\
    &= \frac{1}{2\pi b} \int_{-\infty}^{+\infty} e^{Nh(\xi)}d\xi,
\end{align}
where we defined
\begin{align}
    h(\xi) &\equiv \ln\left[ \frac{\sin(\xi)}{\xi} \right] + i\xi \frac{x}{L}.
\end{align}

To find a solution of this integral, we use a saddle-point approximation or method of steepest descent as explained in Ref.~\onlinecite{yamakawa1971modern}. The saddle-point is at $\xi = \xi_0$, for which
\begin{align}
    h'(\xi_0) = 0.
\end{align}
This approximation then gives rise to~\cite{yamakawa1971modern}
\begin{align}
    P(x) &\approx \frac{1}{2\pi b} \sqrt{\frac{2\pi}{-N h''(\xi_0)}}e^{Nh(\xi_0)}. \label{eq:PxJHtus}
\end{align}
One can show that~\cite{yamakawa1971modern}
\begin{align}
    \xi_0 &= i\mathcal{L}^{-1}\left( \frac{x}{L} \right),
\end{align}
so 
\begin{align}
    h(\xi_0) &= \ln\left( \frac{\sin\left[\mathcal{L}^{-1}\left( \frac{x}{L} \right)\right]}{\mathcal{L}^{-1}\left( \frac{x}{L} \right)} \right) -\frac{x}{L}\mathcal{L}^{-1}\left( \frac{x}{L} \right) .
\end{align}
Furthermore, we have
\begin{align}
    h''(\xi_0) &= \mathcal{L}'\left[ \mathcal{L}^{-1}\left( \frac{x}{L} \right) \right] = \frac{1}{\left(\mathcal{L}^{-1}\right)'\left( \frac{x}{L} \right)} \\
    &= \frac{1}{\sinh^2\left[ \mathcal{L}^{-1}\left( \frac{x}{L} \right)  \right]} - \frac{1}{\left[\mathcal{L}^{-1}\left( \frac{x}{L} \right)\right]^{2}}.
\end{align}
Plugging these in Eq.~\ref{eq:PxJHtus}, we arrive at
\begin{align}
    &P_\text{JH}(x,f=0) = \nonumber \\ &= \frac{ \mathcal{L}^{-1}(x/L)}{\left( 2\pi N b^2 \right)^{1/2}\sqrt{1 - \left[ \mathcal{L}^{-1}(x/L)/\sinh\left( \mathcal{L}^{-1}(x/L) \right) \right]^2}} \times \nonumber \\
    &\times \left[ \frac{\sinh\left( \mathcal{L}^{-1}(x/L) \right)}{\mathcal{L}^{-1}(x/L) e^{x/L\mathcal{L}^{-1}(x/L)}} \right]^N,
\end{align}
as was stated in the main text in Eq.~\ref{eq:JH}. The sadde-point approximation makes no restriction on $x$, but is only valid for $N$ sufficiently large.

\section{Rewriting the exact probability distribution}
\label{app:We}
Here, we will derive the probability density given in Eq.~\ref{eq:WE} of the main text. Introducing the polar angle of joint $j$ along the direction of the force $f$, we can write
\begin{align}
    x &= b\sum_{j=1}^N \cos(\theta_j).
\end{align}
Then, 
\begin{align}
    P(x,f) &= Z^{-1}(f) \int e^{\beta fb\sum_{j=1}^{N}\cos(\theta_j)}\times \nonumber \\ &\times \delta\left( x - b\sum_{j=1}^N \cos(\theta_j) \right) \prod_{j=1}^N \sin(\theta_j) \ d\theta_j d\phi_j,
\end{align}
with $Z(f)$ the partition function of Eq.~\ref{eq:part}. For notational compactness we will express forces in units of $1/(\beta b)$ and lengths in units of $b$ and reinsert those factors at the end of the calculation.

Using the same Fourier transform for the delta function as in Eq.~\ref{eq:FourierDelta}, we can write
\begin{align}
    &Z(f)P(x,f) =  \int_{0}^{\pi} e^{f\sum_{j=1}^{N}\cos(\theta_j)}\times \nonumber \\ &\times \left[ \int_{-\infty}^{+\infty} e^{ik\left( x - \sum_{j=1}^N \cos(\theta_j)\right)}
 \ dk\right] \prod_{j=1}^N  \sin(\theta_j) \ d\theta_j \\
 &=  \int_{-\infty}^{+\infty} e^{ikx}\left[ \int_{0}^{\pi} e^{(f - ik)\cos(\theta)}\sin(\theta) \ d\theta \right]^N \ dk \\
  &= (2\pi)^{N-1} \int_{-\infty}^{+\infty}e^{ikx} \left[ \frac{2\sinh(f - ik)}{f - ik} \right]^N \ dk.
\end{align}
Next, one can use a binomial expansion 
to get
\begin{align}
    \left[ 2\sinh(f - ik) \right]^N &= \left[ e^{f - ik} - e^{-(f - ik)} \right]^N \\
    &= \sum_{s = 0}^{N} \binom{N}{s} (-1)^s e^{-(f - ik)s} e^{(f - ik)(N-s)}.
\end{align}
This allows to rewrite the entire expression as
\begin{widetext}
\begin{align}
Z(f)P(x,f)=(2\pi)^{N-1}  \sum_{s = 0}^{N} \binom{N}{s} (-1)^s  e^{f(N-2s)} 
\int_{-\infty}^{+\infty}\frac{1}{(f-ik)^N}e^{ik(x+2s-N)}\ dk.
\end{align}
The integrand has an $N$-fold pole at $k=-if$. Whenever $x+2s-N$ is positive, we need to close the contour using a $k$ with positive imaginary part so the pole will be outside and the integral is 0. We therefore add a Heaviside factor $\Theta(N-2s-x)$ to only include those contours that contain the pole and use the residue theorem to find

\begin{align}
Z(f)P(x,f) &= \frac{(2\pi)^{N}}{(N-1)!}
e^{fx}
\sum_{s = 0}^{N}
\binom{N}{s}
(-1)^s
(N-x-2s)^{N-1}
\Theta(N-2s-x).
\end{align}
The full probability density with the factors $\beta$ and $b$ reinserted then reads
\begin{align}
P(x, f) &= \left[\frac{\beta fb}{2\sinh(\beta f b)}\right]^N
\frac{e^{\beta f x}}{b(N-1)!}
\sum_{s = 0}^{N}
\binom{N}{s}
(-1)^s
(N-2s-x/b)^{N-1}
\Theta(N-2s-x/b). \label{eq:addb}
\end{align}
In the limit $f\to 0$ this reduces to
\begin{equation}
P_\text{ex.}(x, f=0)=\frac{1}{2^N}
\frac{1}{b(N-1)!}
\sum_{s = 0}^{N}
\binom{N}{s}
(-1)^s
(N-2s-x/b)^{N-1}
\Theta(N-2s-x/b). \label{eq:WEfinal}
\end{equation}
as was stated in the main text in Eq.~\ref{eq:WE}.
\end{widetext}
\end{document}